%% file: Script.tex
\documentclass[conference]{IEEEtran}
\IEEEoverridecommandlockouts
\usepackage[utf8]{inputenc} 
\usepackage[T1]{fontenc}
\usepackage{url}
\usepackage{ifthen}
\usepackage{cite}
\usepackage[cmex10]{amsmath} % Use the [cmex10] option to ensure complicance
\usepackage{textcase}
\usepackage[tablename=Table]{caption}
\usepackage{blindtext}
\usepackage{floatrow}
\usepackage{soul}
\usepackage{changes}
\usepackage{gensymb}
\usepackage{cite}
\usepackage{amsmath,amssymb,amsfonts}
\usepackage{textcomp}
\usepackage{graphicx}
\usepackage{amssymb}
\usepackage{amsfonts}
\usepackage{amsmath}
\usepackage{epsfig}
\usepackage{color}
\usepackage{fancybox}
\usepackage{textcomp}
\usepackage{multirow}
\usepackage{setspa ce}
\usepackage{psfrag}
\usepackage{booktabs}
\usepackage{float}
\usepackage{algorithm}
\usepackage{algpseudocode}
\usepackage{mathtools, nccmath, bigints, amsfonts}
\usepackage{array}

\DeclarePairedDelimiter\floor{\lfloor}{\rfloor}

\usepackage{caption}
\usepackage{subcaption}
\usepackage{mathrsfs}							

\usepackage{placeins}

\newfloatcommand{capbtabbox}{table}[][0.4\textwidth]

\IEEEaftertitletext{\vspace{-2\baselineskip}}
\makeatletter 
\newcommand\citecolor[1]{\@namedef{keycolor#1}{\color{blue}}}
\makeatother
\input macro
\def\BibTeX{{\rm B\kern-.05em{\sc i\kern-.025em b}\kern-.08em
    T\kern-.1667em\lower.7ex\hbox{E}\kern-.125emX}}
\begin{document}

\title{Bayesian Neural Network Detector for an Orthogonal Time Frequency Space Modulation} 

 \author{%
   \IEEEauthorblockN{
   					Alva Kosasih\IEEEauthorrefmark{1},
   					 Xinwei Qu\IEEEauthorrefmark{1},
                     Wibowo Hardjawana\IEEEauthorrefmark{1}, 
                    Chentao Yue\IEEEauthorrefmark{1},
                    and Branka Vucetic\IEEEauthorrefmark{1}\\
                    }
   \IEEEauthorblockA{\IEEEauthorrefmark{1}%
                     Centre of Excellence in Telecommunications, University of Sydney, Sydney, Australia.  \\
                    \{alva.kosasih,wibowo.hardjawana,chentao.yue,branka.vucetic\}@sydney.edu.au,
                     xiqu4217@uni.sydney.edu.au.   }
                    
   %\IEEEauthorblockA{\IEEEauthorrefmark{2}%
                    % Mobilizing Information Technology Lab., National Taiwan University of Science and Technology, Taipei, Taiwan. }
 }

\maketitle

%%%%%%
%% Abstract: 
%% If your paper is eligible for the student paper award, please add
%% the comment "THIS PAPER IS ELIGIBLE FOR THE STUDENT PAPER
%% AWARD." as a first line in the abstract. 
%% For the final version of the accepted paper, please do not forget
%% to remove this comment!
%%
\begin{abstract}
% Abstract goes here.
The orthogonal time frequency space (OTFS) modulation is proposed for beyond 5G wireless systems to deal with high mobility communications. The existing low complexity OTFS detectors' performance is suboptimal in rich scattering environments where there are a large number of moving reflectors that reflect the transmitted signal towards the receiver.  In this paper, we propose an OTFS detector, referred to as the BPICNet OTFS detector that integrates NN, Bayesian inference, and parallel interference cancellation concepts. Simulation results show that the proposed  detector  outperforms  the state-of-the-art.
\end{abstract}


%% The paper must be self-contained. However, if you are referring to
%% a full version for checking certain proofs, please provide the
%% publically accessible location below.  If the paper is completely
%% self-contained, you can remove the following line from your
%% submission.

\section{Introduction}

\begin{figure}
\centering
\includegraphics[scale=0.45]{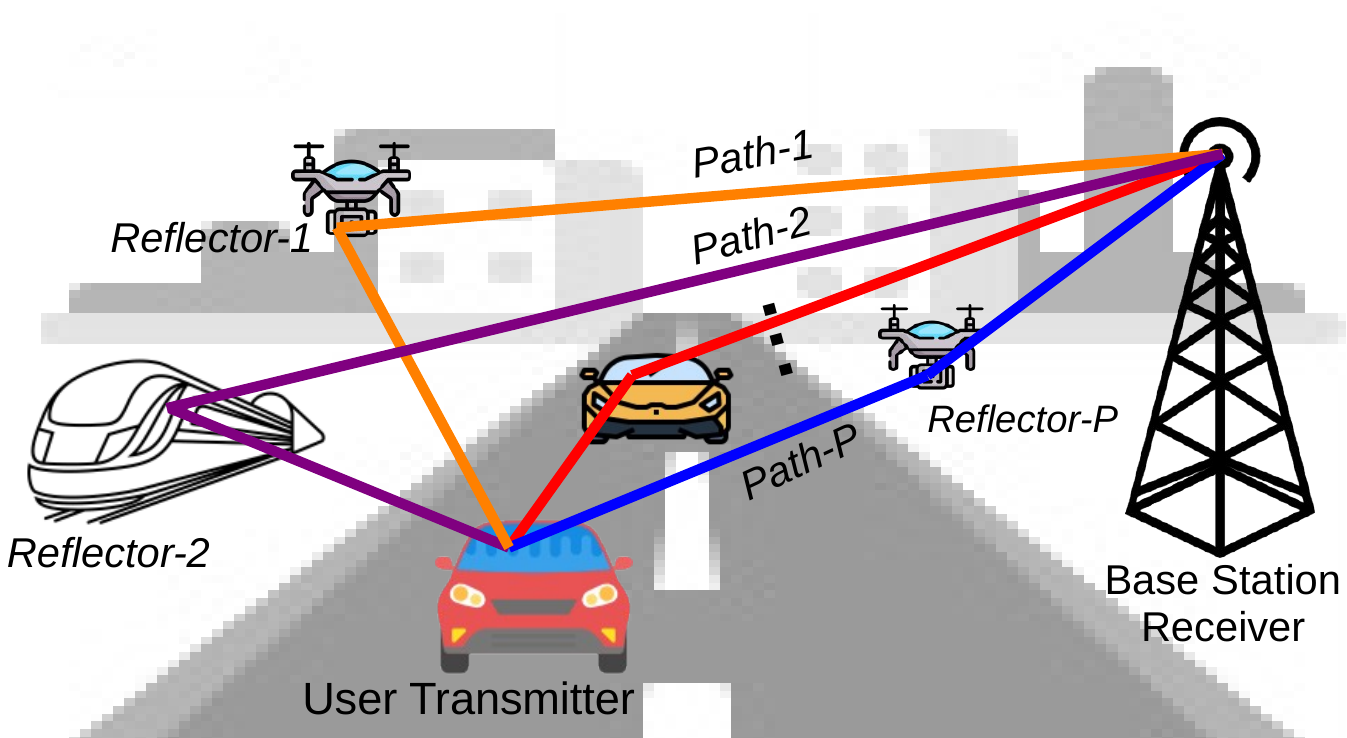}
\caption{High mobility environment}
\label{fig:friendly}
\end{figure}

Beyond 5G wireless system enables high-speed communications which involve unmanned aerial vehicles, trains, and autonomous cars \cite{hashimoto2020channel}.
As illustrated in Fig. \ref{fig:friendly}, there is a high speed transmitter, several reflectors (i.e. the other moving vehicles/objects around the vicinity of the transmitter), and a base station receiver in an uplink scenario. The moving reflectors generate  a high inter-carrier interference. This causes significant performance degradation in the current OFDM modulation \cite{raviteja2018interference,OTFS_2021}.
 To address this issue, an orthogonal time frequency space (OTFS) modulation  that multiplexes the information symbols in the delay-Doppler (DD) domain was proposed in \cite{Hadani2017}. The OTFS performance highly depends on its symbol detector at the receivers.
%  An OTFS system is equipped with a symbol detector to recover the transmitted symbols in the presence of the ICI and ISI. 
%  Therefore, a high performance OTFS detector ensures an excellent performance in high mobility communications. 
To date, there are many works on OTFS symbol detectors. They can be categorised into two groups  i.e., classical \cite{singh2020low,raviteja2018interference,Luping2021TVT,yuan2020iterative,FLONG2022EP-OTFS,Akosasih2021otfs} and neural-network (NN) \cite{Naikoti2021VTC,Enku2021CNN} based OTFS  detectors.

A classical minimum-mean-square-error (MMSE) OTFS detector \cite{singh2020low} provides a low complexity but demonstrating a suboptimal performance. A classical iterative message passing (MP)  detector for OTFS systems  \cite{raviteja2018interference} performs much better than the MMSE OTFS detector, where the 
 interference is approximated using Gaussian functions.
 % and mitigated it from the received signal to yield the symbol estimates. 
A variational Bayes (VB) OTFS detector has also been considered in \cite{Varitional_Bayes} to improve the convergence of the MP OTFS detector.
Unfortunately, their performance degrades significantly  when the number of moving reflectors rises, referred to as a rich scattering environment. 
To achieve excellent performance under a rich scattering environment, the expectation propagation (EP) \cite{FLONG2022EP-OTFS} OTFS detectors have been proposed. However, EP needs to perform matrix inverse operation/approximation in each iteration, resulting in a high complexity. 
To avoid the high complexity issue, a unitary approximate MP (UAMP) \cite{yuan2020iterative} and a Bayesian parallel interference cancellation (BPIC) \cite{Akosasih2021otfs}  OTFS detectors were proposed. The UAMP applies singular value decomposition (SVD) on the OTFS channel matrix, transforms the received signal using the unitary matrix, and then performs the AMP scheme based on the transformed received signal to yield the symbol estimates. The BPIC estimates the symbols by iteratively mitigating the interference from the received signal in a parallel manner. The UAMP and BPIC demonstrate similar performance. However, there exists a performance gap between UAMP and BPIC with EP.
% They avoid performing multiple matrix inverse operations.
% using the parallel interference cancellation (PIC) scheme and achieves a close EP OTFS performance. 
% However, their performance is still suboptimal. 
% \gray{under a rich scattering environment.} 

The second category of OTFS detectors, the NN based OTFS detectors have a  better detection performance compared to the classical OTFS detectors. 
% \gray{However, their complexity is high due to a large number of NN parameters.} 
In \cite{Naikoti2021VTC,Enku2021CNN}, the NNs are trained using simulated data consisting of the received signals and transmitted symbols to approximate the detection mapping function from the received signals to the transmitted symbol estimates. Unfortunately, the numbers of trainable NN parameters in both NN detectors are very large. This leads to high complexity and causes the parameters difficult to be optimized. 
{A model-driven NN technique that introduces a few trainable NN parameters for classical iterative algorithms has been proposed in \cite{2020_HHe_TSP_OAMPNet,EPANet,2020_Khani_TWC_Adaptive}.
% To avoid using a large number of trainable NN parameters, a model-driven NN technique, where a few of NN parameters is inserted in each iteration of the classical iterative algorithm has been proposed in \cite{2020_HHe_TSP_OAMPNet,EPANet,2020_Khani_TWC_Adaptive}. 
 The OAMPNet \cite{2020_HHe_TSP_OAMPNet} and the EPANet \cite{EPANet} add at most four NN parameters in each iteration of the orthogonal AMP (OAMP) and approximated EP detectors, respectively. However, they suffer from high complexity since they perform matrix inverse operations/approximations in every iteration. A model-driven NN detector named MMNet in  \cite{2020_Khani_TWC_Adaptive} can achieve at par OAMPNet performance without performing matrix inverse operations. However, it requires a large number of additional NN parameters compared to the OAMPNet and EPANet, and thus its complexity is high. Note that all the above model-driven NN detectors are implemented in multiple-input multiple-output systems. To the best of our knowledge, there is no prior work on the model-driven NN detectors in the OTFS systems.}

\begin{figure*}
\centering
\includegraphics[width=0.76\textwidth]{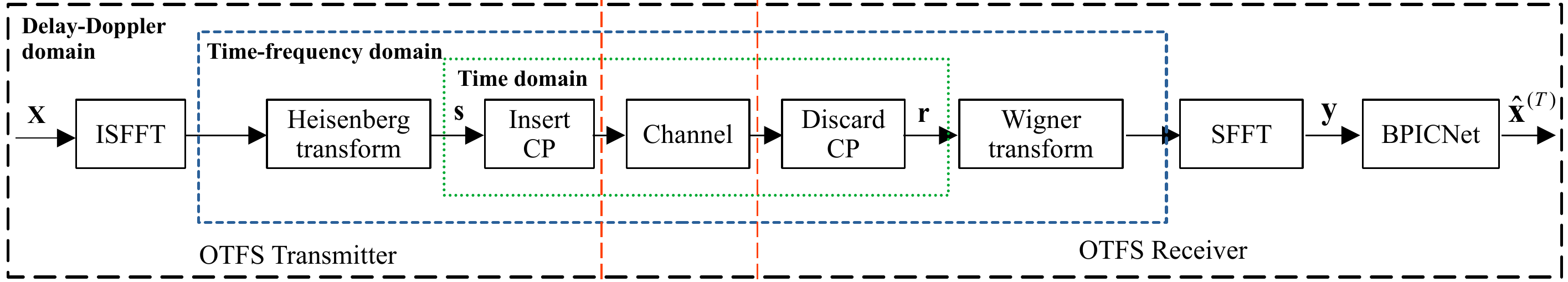}
\caption{The system model of OTFS modulation scheme}
\label{fig:sys-mod-general}
\end{figure*} 
In this paper, we propose a low complexity NN based OTFS detector that can achieve a high detection performance under rich scattering environments. This is realized  by integrating the BPIC and NN which we refer to as  the BPIC Network (BPICNet) detector. 
The original BPIC consists of three modules \cite{kosasih2020linear}, namely: 1) the Bayesian symbol observation (BSO) module which employs a parallel interference cancellation (PIC) scheme to calculate the mean and variance of the transmitted symbols; 2) the Bayesian symbol estimate  (BSE) module  which calculates the soft symbol estimates based on the BSO outputs; 3) the decision statistic combining (DSC) module which weights the soft symbol estimates and their variance in the previous and current  iterations.
We develop the BPICNet by replacing the linear function\footnote{A standard NN linear function is $\theta \mathcal{X} +b$, where $\theta$ is an NN parameter, $\mathcal{X}$ is an input, and $b$ is a bias parameter.}  in the standard NN's neurons with modified BPIC functions, at each layer.
These functions except the one in the BSE  are treated as neurons where in each of the neurons there is a tunable parameter, referred to as the NN parameter. More specifically, two neurons are employed in the BSO  while a neuron is employed in the DSC. The Bayesian symbol estimate (BSE) block contains a function calculating the soft symbol estimates and variances based on the Bayesian inference.

The main contribution of this paper is the first development of an OTFS detector that integrates NN, Bayesian inference, and PIC concepts. This results in a significant detection improvement as compared to its predecessor, BPIC detector \cite{Akosasih2021otfs}, as well as the other low complexity OTFS detectors \cite{singh2020low,raviteja2018interference,Luping2021TVT,yuan2020iterative}. 
Furthermore, unlike the state-of-the-art NN based OTFS detectors \cite{Naikoti2021VTC,Enku2021CNN} which optimize a large number of NN parameters, the BPICNet OTFS detector employs only three NN parameters in each layer and thus the parameters are much easier to be optimized. 
    % The simulation results confirm that the BPICNet detector can achieve a similar performance with the current best OTFS detector i.e., the EP detector which has much higher complexity compared to the BPICNet due to performing matrix inverse operation in every iteration.

%Note that we follow the OTFS model, considered in \cite{raviteja2018practical}, where the cyclic prefix (CP) is only inserted in the beginning of the OTFS frame,  regarded as the most efficient OTFS model in terms of latency and spectral efficiency.
%The main contributions are summarized as follows:
%We aim to significantly improve the performance of the current state-of-the-art MMSE based detectors with the same computational complexity order.
% The main contribution of this paper is the development of the B-PIC-DSC detector for an OTFS system  that can achieve a high BER performance, in the presence of a strong ICI due to a large number of moving reflectors. The simulation results demonstrate that our proposed detector is able to achieve a BER of $10^{-5}$, in the presence of a large number of high mobile reflectors. This is in contrast to other OTFS detectors that fail to work in such an environment. 

{\bf Notations}: $a$, $\qa$ and $\qA$ denote scalar, vector, and matrix respectively. $\mathbb{C}^{M\times N}$ denotes the set of $M\times N$ dimensional complex matrices. We use $\qI_N$, $\qF_N$, and $\qF_N^{H}$ to represent an $N$-dimensional identity matrix, $N$-points discrete Fourier Transform (DFT) matrix, and $N$-points inverse discrete Fourier transform (IDFT) matrix. $(\cdot)^\top$
% $(\cdot)^{H}$, $(\cdot)^*$, and  $[\cdot]_M$ 
represent the transpose
% , Hermitian, and  conjugate,  and mod-$M$ 
operation. We define $\qa = {\sf vec}(\qA)$ as the column-wise vectorization of matrix $\qA$ and its inverse operation $\qA = {\sf vec}^{-1}(\qa)$.
%Notation $diag(\textbf{a})$ denotes a diagonal matrix whose diagonal is the vector $\textbf{a}$, $diag(\qA)$ denotes a vector that is the diagonal of the matrix $\textbf{A}$ and $diag_0(\qA)$ denotes the operation that forces the diagonal of matrix $\qA$ to zero. 
The Kronecker product is denoted as $\otimes$. 
% , and $\qa \cdot \qb$, $\qa ./ \qb$ and $|\cdot|^2$ denote the element-wise production production, division and magnitude squared operation respectively. 
% We use $\bf 0$ and $\bf 1$ to represent the adequately long vectors full of 0 and 1 respectively. The superscript $(\cdot)^{(t)}$ denotes the $t$-th iteration and 
The Euclidean distance of vector $\qx$  is denoted as $\|\qx\|$.
% $[K] = \{1,2, \cdots, K \}$ is the set of all natural numbers up to $K$.
We use $\mathcal{N}(x: \mu, \sigma)$ to express a single variate Gaussian distribution of a random variable $x$ where $\mu$ is the mean and $\sigma$ is the variance.

% \begin{figure}
% \centering
% \subfloat[Before adding CP]
% {\includegraphics[scale=0.12]{otfs_tf.pdf}}\hfill
% \centering
% \subfloat[After adding CP]
% {\includegraphics[scale=0.16]{otfs_time_domain.pdf}}
% \caption{An OTFS frame structure}
% \label{fig:otfs-frame}
% \end{figure}

\begin{figure*}
\centering
\includegraphics[width=0.89\textwidth]{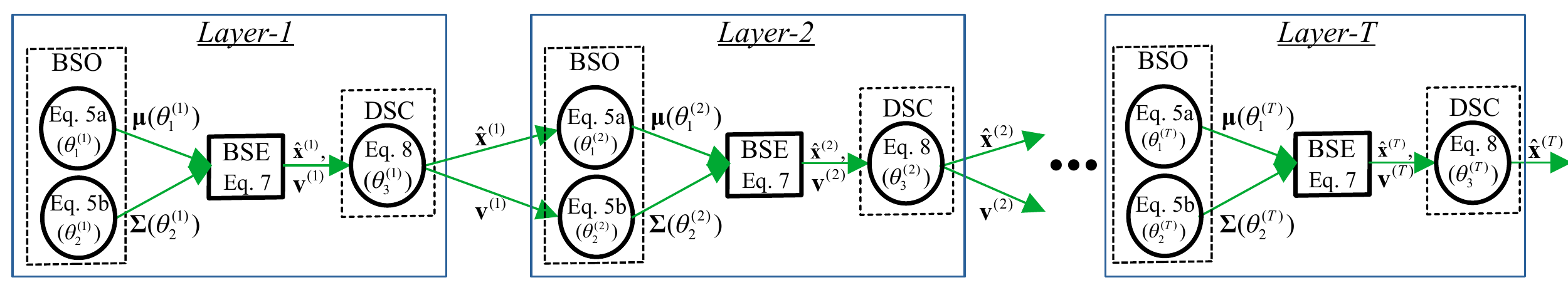}
\caption{The BPICNet OTFS detector with $\theta_1^{(t)}, \theta_2^{(t)}, \text{ and } \theta_3^{(t)}, t\in \{1,\dots,T\},$ as the NN parameters} 
\label{fig:BPICNet}
\end{figure*}

\input{syst_mod}

\section{BPICNet OTFS Detector}

In this section, we describe the development of the  B-PICNet detector for an OTFS system. There are $T$ layers, where in each layer there are three neurons represented by the circles and a BSE block, as illustrated in Fig. \ref{fig:BPICNet}. 
% The term layer is used to emphasize that we embed the BPIC into NN. Specifically, we replace a standard linear function in the NN's neurons  by the modified BPIC functions. 
Since the BPICNet embeds the BPIC into NN, it can be described by the three BPIC modules i.e.,  the BSO, BSE, and DSC modules. 
% Note that the NN parameters in the BPICNet are trained using backpropagation to minimize the mean square error of the symbol estimates.
% Therefore, in each BPICNet layer, there exist three modules i.e., the BSO, BSE, and DSC modules similarly to its predecessor the BPIC. 
% \red{Specifically, inside of each BPICNet layer, there are two neurons computing BSO functions of the BPIC, a BSE function that modules i.e., the BSO, BSE, and DSC modules.}
We omit the subscript $\rm eff$ for notational simplicity.

\subsection{Bayesian Symbol Observation (BSO)}

The posterior probability distribution of the transmitted symbols $\qx = [x_1, \cdots, x_q, \cdots, x_{KL}]$ in \eqref{eq:sysddmod-tx-heisenberg} given the received signal $\qy$ and channel matrix $\qH$ in \eqref{sysddmod-y=hx+w} can be approximated by using a product of independent Gaussian functions as $p(\qx|\qy)  \approx \prod_{q=1}^{KL} \mathcal{N}\left( x_q: \mu_{q}(\theta^{(t)}_1), \Sigma_q(\theta^{(t)}_2) \right)$\cite{kosasih2020linear}.  
$ \mu_{q}(\theta^{(t)}_1)$ and $\Sigma_q(\theta^{(t)}_2)$  are the mean and variance of the $q$-th Gaussian function of $x_q$ in  $t$-th layer which can be obtained from  the matched filter based PIC scheme, given as
% \footnote{We manipulate the expression of the mean of the BSO module in \cite{kosasih2020linear} to integrate it with the neural network parameter $ \theta_1^{(t)}$. The detail of the manipulation is not given in here due to space limitation.}
\begin{subequations}\label{eA1_a0102}
\begin{equation}\label{eA1_a02}
\mu_{q}(\theta^{(t)}_1) = \hat{x}_{q}^{(t-1)} + \theta_1^{(t)} \frac{\qh_q^{\top} }{\| \qh_q \|^2} \left( \qy-\qH\hat{\qx}^{(t-1)} \right),
\end{equation} 
\begin{equation}\label{eA1_a01}
\Sigma_q(\theta^{(t)}_2) =  \frac{\theta_2^{(t)}}{\left(\qh_q^\top \qh_q  \right)^2} \left( \sum_{\substack{j=1 \\  j\neq q}}^{KL} \left(\qh_q^\top \qh_j  \right)^2 v_j^{(t-1)} +   \left(\qh_q^\top \qh_q   \right) \sigma^2 \right).
\end{equation}
\end{subequations}
$\qh_q$ is the $q$-th column of matrix $\qH$,  $\hat{\qx}^{(t-1)} $
is a vector of symbol estimates in layer $t-1$, $v_j^{(t-1)}$ is the $j$-th element in a vector variance  of symbol estimates   $\qv^{(t-1)}$,    $ \theta_1^{(t)} $ is an NN parameter used to adjust the residual error of the BSO symbol estimates in layer $t$, and $ \theta_2^{(t)}$ is an NN parameter  used to weight the BSO variance in layer $t$.
Note that we use the MMSE scheme to produce the initial value of the symbol estimates,
\begin{equation}\label{MMSE_init}
\hat{\qx}^{(0)}= \left(\qH^{\top} \qH + \sigma^2 \qI_{KL} \right)^{-1}\qH^{\top} \qy.
\end{equation}
% The variance $ \Sigma_q $ is given as\cite{kosasih2020linear}
% \begin{equation}\label{eA1_a01}
% \Sigma_q^{(t)} =  \frac{\theta_2^{(t)}}{\left(\qh_q^\top \qh_q  \right)^2} \left( \sum_{\substack{j=1 \\  j\neq q}}^{KL} \left(\qh_q^\top \qh_j  \right)^2 v_j^{(t-1)} +   \left(\qh_q^\top \qh_q   \right) \sigma^2 \right),
% \end{equation}
The vectors $\qmu(\theta^{(t)}_1) = \left[\mu_1(\theta^{(t)}_1), \dots, \mu_{KL}(\theta^{(t)}_1) \right]^{\top} $ and  $\qSigma(\theta^{(t)}_2) = \left[\Sigma_1(\theta^{(t)}_2), \dots, \Sigma_{KL}(\theta^{(t)}_2) \right]^{\top} $ are then  then  forwarded to the BSE module, as depicted in Fig. \ref{fig:BPICNet}.

\subsection{Bayesian Symbol Estimator (BSE)}
      
In the BSE module, we compute the Bayesian symbol estimates and their variances based on the outputs of the BSO module.  
% Based on the factorization of $p(\qx|\qy)$  in \eqref{iid_assumption}, we infer the symbol estimate $\hat{x}_q^{(t)}$ by using the maximum a posteriori criterion, given as 
% \begin{flalign}\label{MAP_component}
% \hat{x}_q^{(t)} &= \arg \max_{a \in \Omega} \hat{p}^{(t)}(x_q=a|\qy).
% \end{flalign}
% Note that the MAP criterion in \eqref{MAP_component}  has a linear complexity since the inference is performed for each $q$-th symbol estimate. 
The Bayesian symbol estimate and variance of the $q$-th transmitted symbol are respectively given as
\begin{subequations}\label{eA1_b0102}
\begin{equation}\label{eA1_b01}
\hat{x}_q^{(t)} =\Ex\left[x_q \Big| \mu_q(\theta^{(t)}_1) ,\Sigma_q(\theta^{(t)}_2) \right] =\sum_{a \in \Omega} a p^{(t)}\left(x_q=a|\qy\right),
\end{equation}
\begin{equation}\label{eA1_b02}
v_q^{(t)}=\Ex  \left[ \left| x_q  - \Ex\left[x_q \Big|  \mu_q(\theta^{(t)}_1) ,\Sigma_q(\theta^{(t)}_2) \right] \right|^{2} \right],
\end{equation}
\end{subequations}
where $p^{(t)}\left(x_q|\qy\right) =  {\cal{N}}\left( x_q:\mu_q(\theta^{(t)}_1), {\Sigma}_q(\theta^{(t)}_2) \right)$ is obtained from the BSO module 
% The Gaussian function in \eqref{eA1_b01} is normalized, so that the sum of the total probability is unity. 
and it is normalized so that  $\sum_{a\in \Omega} p^{(t)}\left(x_q=a|\qy\right) =1$.
The output vectors of the BSE module, $\hat{\qx}^{(t)} = \left[\hat{x}_1^{(t)} , \dots, \hat{x}_{KL}^{(t)}  \right]^\top$ and $\qv^{(t)} = \left[v_1^{(t)} , \dots, v_{KL}^{(t)}  \right]^\top$  are then sent to the DSC module.

\subsection{Decision Statistics Combining (DSC)}

In the DSC module, we linearly combine the symbol estimate  and variance in the subsequent layers, i.e.,
% The correlation between $\hat{x}_q^{(t)}$ and $\hat{x}_q^{(t-1)}$ in the B-PIC-DSC detector is low in the early iteration stages\cite{kosasih2020linear}. 
% Such a feature can be exploited to increase the diversity of symbol estimates by forming decision statistics,  referred to as the DSC concept. The decision statistics consist of a linear combination of the symbol estimates in two consecutive iterations 
\begin{subequations} \label{DSC_complete}
\begin{equation}\label{DSC}
\hat{x}_{q}^{(t)} \leftarrow \left( 1-\rho_{q}(\theta_3^{(t)}) \right)    \cdot  \hat{x}_q^{(t-1)}  +   \rho_{q}(\theta_3^{(t)})  \cdot  \hat{x}_q^{(t)},
\end{equation}
\begin{equation}\label{DSC_Var}
v_{q}^{(t)} \leftarrow  \left( 1-\rho_{q}(\theta_3^{(t)}) \right)  \cdot   v_q^{(t-1)}  +   \rho_{q}(\theta_3^{(t)}) \cdot v_q^{(t)},
\end{equation}
% \begin{equation}\label{DSC_coeff}
% \end{equation}
\end{subequations}
where $ \rho_{q}(\theta_3^{(t)}) \triangleq  \frac{e_q^{(t-1)}}{e_q^{(t)}+e_q^{(t-1)}},$ $ e_q^{(t)}  = \theta_3^{(t)} \left| \qh_q^{\top} \left(  \qy - \qH \hat{\qx}^{(t)} \right) \right|^2, $  and $ \theta_3^{(t)}$ is an NN parameter used to adjust the square error of the symbol estimates in layer-$t$.  Note that $\theta_3^{(t)}$ allows the BPICNet to fine-tune the DSC coefficients and thus regulates the update portion of the symbol estimates and variances in each layer. The weighted symbol estimate and variance vectors, $\hat{\qx}^{(t)} $ and $\qv^{(t)} $, are then fedback to the BSO module to proceed the computations in the next layer. The processes are repeated for $T$ times as depicted in Fig. \ref{fig:BPICNet}. The complete pseudo-code of the BPICNet OTFS detector is given in Algorithm \ref{A1}.

% The iterative process is terminated if the following condition is satisfied,
% \begin{equation}\label{eq_convergence}
%  \|x_{{\rm DSC},q}^{(t)} - x_{{\rm DSC},q}^{(t-1)} \| \leq \zeta  \text{  or  } t = t_{\rm max}, 
% \end{equation}
% where  $\zeta$ is the minimum acceptable difference of $x^{(t)}_{{\rm DSC},q}$ in two consecutive iterations, and $t_{\rm max}$ is the maximum number of iterations. We then use $x_{{\rm DSC},q}^{(t)} $ as the input of the BSO module by assigning the value of $x_{{\rm DSC},q}^{(t)}$ to $x_{{\rm PIC},q}^{(t)}$,
% \begin{equation}\label{assign}
% x_{{\rm PIC},q}^{(t)} \leftarrow x_{{\rm DSC},q}^{(t)}.
% \end{equation}
 %\gray{,\begin{equation}\label{Assign}x_{{\rm PIC},q}^{(t)} \leftarrow  {x}_{{\rm DSC},q}^{(t)},  \text{ and }  v_q^{(t)} \leftarrow v_{{\rm DSC},q}^{(t)},  q =1,\dots,KL.\end{equation}}
%The iteration is repeated until \eqref{eq_convergence} is zsatisfied.
%Then, the hard decision is made according to}
%\begin{equation}\label{Hard_Dec}
%x_{q}^{\rm hard} =\arg \min_{x_q \in \Omega} \| x_q - x_{{\rm PIC},q}^{(\top)}  \|^2, q =1,\dots,KL
%\end{equation}

\begin{algorithm}
\caption{The BPICNet OTFS detector}
\label{A1}
\begin{algorithmic}[1]
\State {\textbf{Input: }$\qy, \qH, \sigma^2, T $}
	\For {$t=1,\dots, T$}
% 		\For {$q=1,\dots, KL$ (parallel execution)}
	    		\If {$t=1$}
			\State {Compute \eqref{MMSE_init}}
			\EndIf
				% \For {$q=1,\dots, KL$} (Parallel Execution)
	    		\Statex \textbf{\quad The BSO Module:}
    			\State \quad Compute \eqref{eA1_a02} to obtain $\qmu(\theta^{(t)}_1)$
    			\State \quad Compute  \eqref{eA1_a01} to obtain  ${\qSigma}(\theta^{(t)}_2) $
    			\Statex \textbf{\quad The BSE Module:}
    			\State \quad Compute \eqref{eA1_b01} to obtain $\hat{\qx}^{(t)} $ 
    			\State  \quad Compute \eqref{eA1_b02} to obtain $\qv^{(t)} $
    			\Statex \textbf{\quad The DSC Module:}
    % 			\State Compute $e_{q}^{(t)}, q \in [KL] $  in \eqref{DSC_error} 
    % 			\State Compute  $\rho_{q}^{(t)}, q \in [KL] $  in \eqref{DSC_coef} 
    			\State \quad Compute \eqref{DSC} to obtain weighted $\hat{\qx}^{(t)} $ 
    			\State \quad Compute  \eqref{DSC_Var}  to obtain weighted $\qv^{(t)} $
    % 			\EndFor
    			%\State Compute $v_{q}^{(t)}$ in \eqref{DSC_Var}  
    % 			\State Compute \eqref{assign}  
    % 		\EndFor
    % 		\If {$ \|x_{{\rm DSC},q}^{(t)} - x_{{\rm DSC},q}^{(t-1)} \| \leq 10^{-4} $}
    % 			\State {break}
    % 		\EndIf
	\EndFor
\State {\textbf{Output: }  $\hat{\qx}^{(T)} = \left[\hat{x}_{1}^{(T)}, \cdots, \hat{x}_{KL}^{(T)}  \right]^\top$}
% 		\State $t_{\rm last}\leftarrow t$ 
%\State Calculate hard symbol estimates from $\hat{\qx}^{(T)}$
%\State \textbf{Return:}
\end{algorithmic}
\end{algorithm}

\newcolumntype{L}[1]{>{\raggedright\let\newline\\\arraybackslash\hspace{0pt}}m{#1}}
\newcolumntype{C}[1]{>{\centering\let\newline\\\arraybackslash\hspace{0pt}}m{#1}}
\newcolumntype{R}[1]{>{\raggedleft\let\newline\\\arraybackslash\hspace{0pt}}m{#1}}

\begin{table}\small
  \begin{tabular}{| C{1.4cm} | C{3.cm}  | C{0.4cm} | C{2.5cm} |}
  \hline
 OTFS detector 	   &    Complexity	&  {T} &  Numerical example $K=7,L=12,M=4,P=14,$ $(\times 10^{5})$	\\ 
     	 \hline
  % \hline
  %   		MMSE \cite{singh2020low}			 &  		$\mathcal{O} ((KL)^3)$ 	 	& \blue{1} &	$5.9270$		\\
  \hline
    		MP 	\cite{raviteja2018interference}		 & 	$\mathcal{O} ((KL)^2PMT)$ 		& {9} & 			{$35.5622$	}		\\
     	 \hline
    		UAMP	\cite{yuan2020iterative}	 & 		$\mathcal{O}((KL)^3 +(KL)^2T)$ 	 		& {9} & 		{$6.56208$}		\\
    		\hline
    		BPIC \cite{Akosasih2021otfs}			 &		$\mathcal{O}((KL)^3 + (KL)^2T)$	 	& {8} & 	{$6.4915$	}		\\
    		\hline
     	BPICNet 	 &  		$\mathcal{O}((KL)^3 + KL+ (KL)^2T )$ 	& 	 {9} &  	{$6.56208$	}		\\
     	 \hline
     	EP 	\cite{FLONG2022EP-OTFS}	 &	$\mathcal{O}((KL)^3T)$ 	 &	 {5} & {$29.6352$ }			\\
     	 \hline
  \end{tabular}  \label{T1}
  \caption{Computational complexity comparison}
\end{table}

\begin{figure*}
\centering
\subfloat[SER vs number of paths, $L=12$, $l_{\rm max}=11$]
{\includegraphics[scale=0.278]{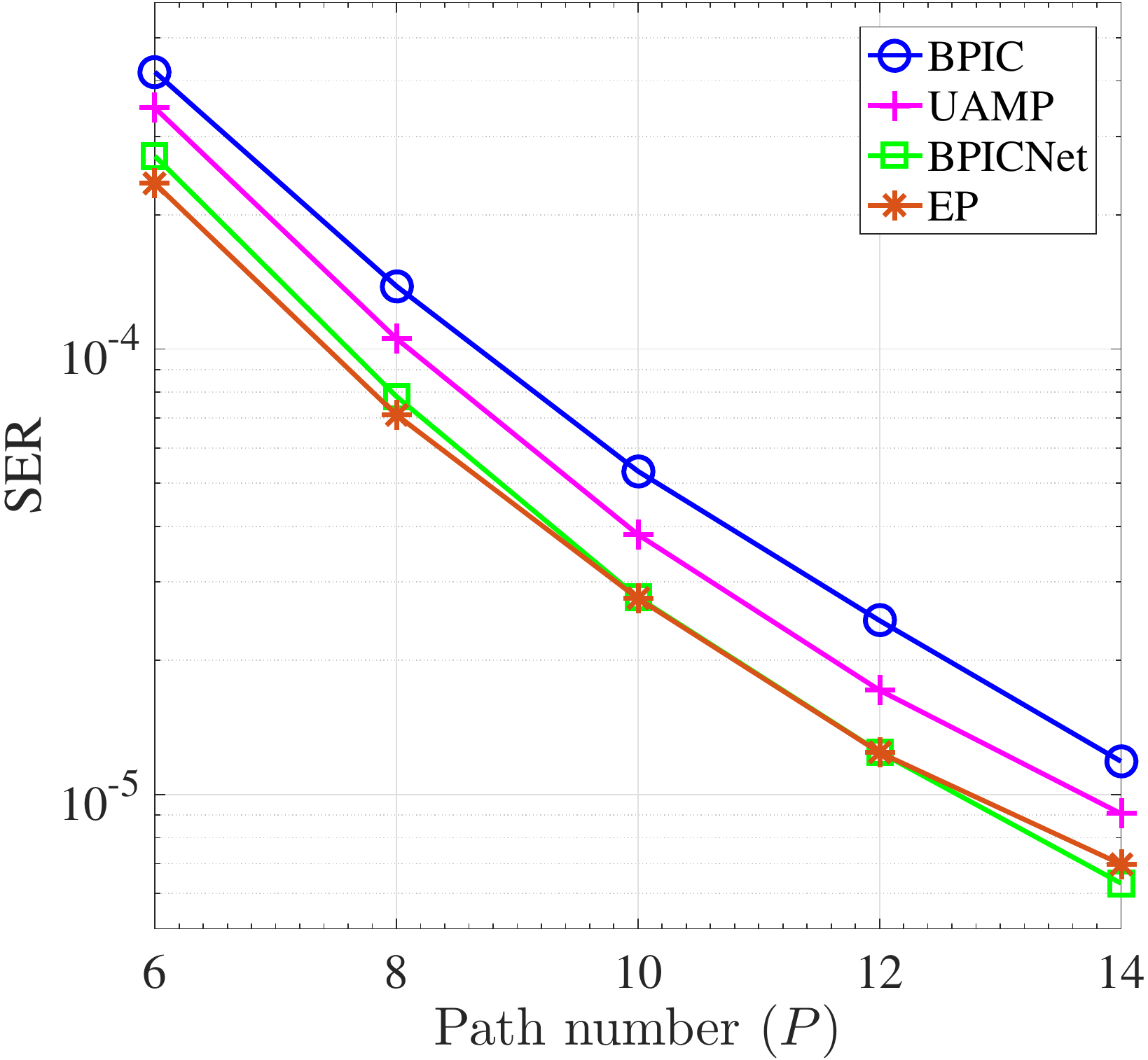}} \quad\quad\quad\quad
\subfloat[SER vs OTFS matrix size, $P=6$, $l_{\rm max}=L-1$]
{\includegraphics[scale=0.298]{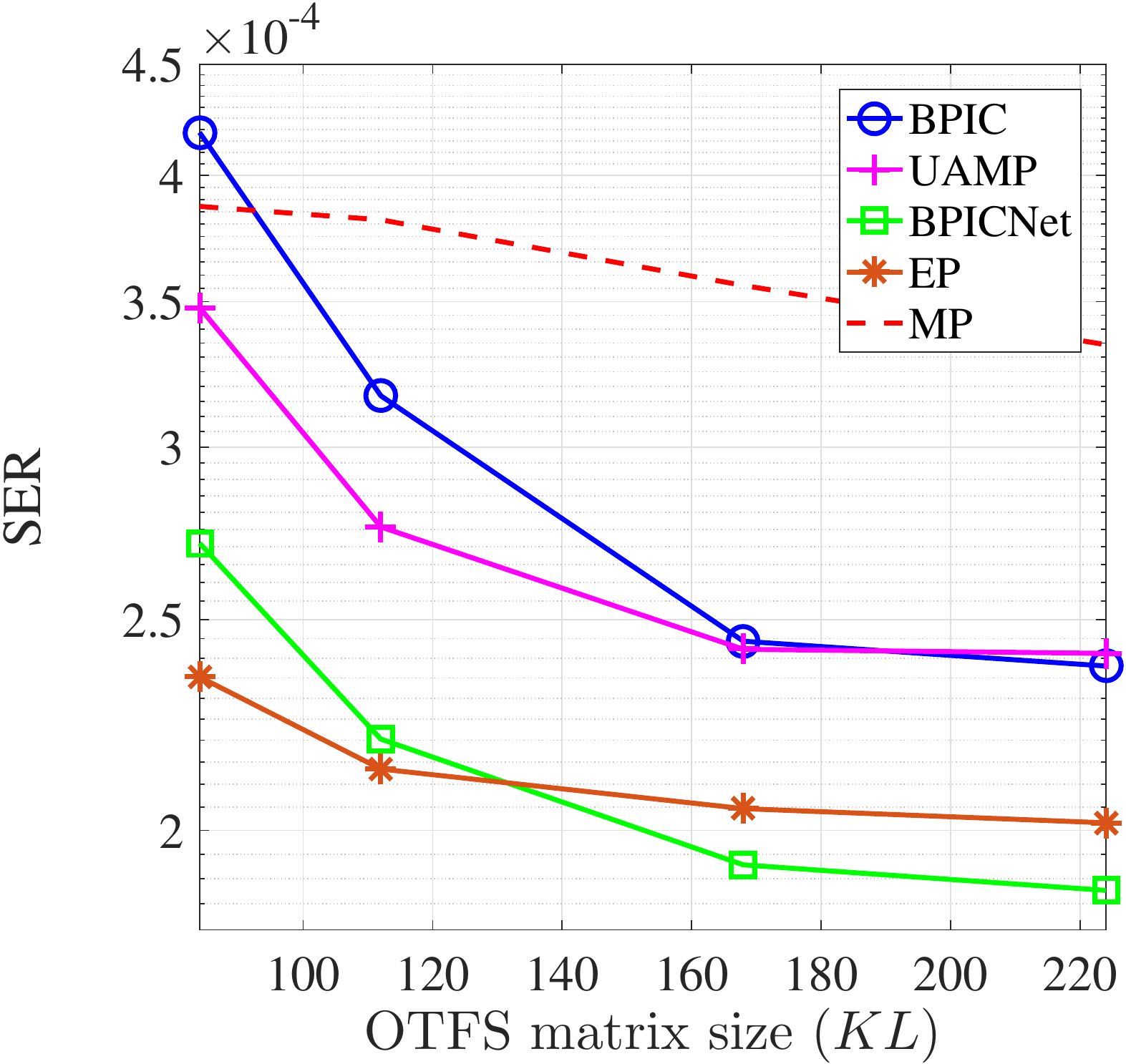}} \quad\quad\quad\quad
\subfloat[SER vs iterations, $P=14$, $L=12$, $l_{\rm max}=11$]
{\includegraphics[scale=0.268]{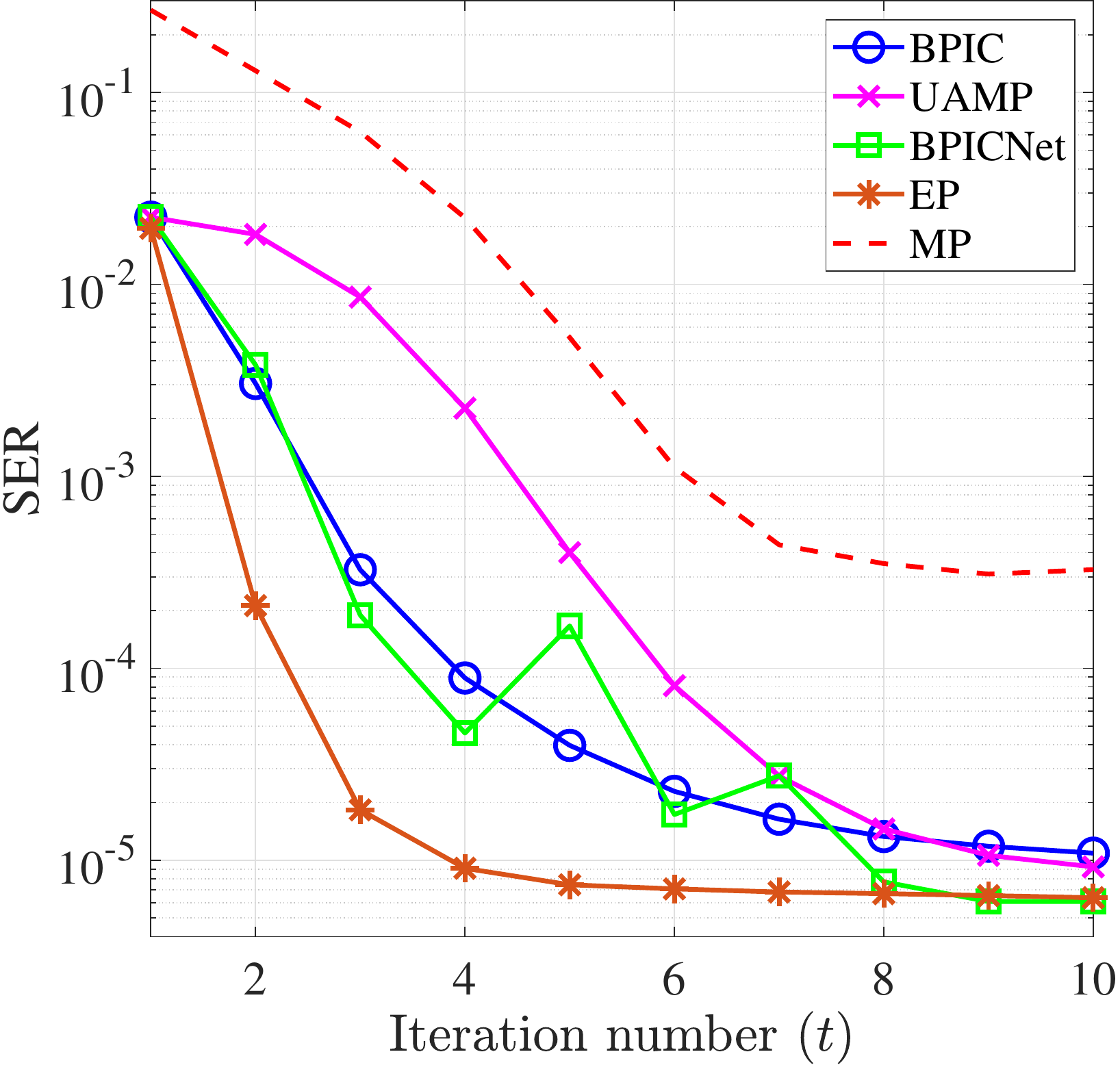}}
\caption{The SER performance comparison}
\label{Fig:SER}
\end{figure*}

\section{{Complexity Analysis}}\label{Complexity}

In this section, we analyze the complexity of the proposed BPICNet OTFS detector. Algorithm 1 specifies that the BPICNet performs matrix vector multiplications  at each layer in  \eqref{eA1_a0102}, \eqref{eA1_b0102}-\eqref{DSC_complete} and therefore the cost is $\mathcal{O} ((KL)^2T)$ in terms of the number of multiplications. The matrix inverse operation, performed in  \eqref{MMSE_init}  requires $\mathcal{O}((KL)^3 )$.  Note that the complexity of multiplying three NN parameters in each layer is small as its cost is only $\mathcal{O}(KL)$. As given in Table 1, {the complexity of the BPICNet is around four times lower than the EP's complexity and similar to that of the BPIC and UAMP.}
% Although the proposed B-PIC-DSC OTFS detector has a higher complexity compared to the other detectors, this comes with a significant BER performance gain in the presence of strong ICI, resulted from the presence of a large number of reflectors, as will be discussed in the next section. 
%Besides, \blue{$KL=56$ in the smallest LTE resource block usage considered in our experiment\cite{surabhi2019diversity}}.
%The B-PIC-DSC OTFS detector requires more computing power to achieve a high detection performance, especially in the rich scattering environment.

\section{Simulation Results}\label{Sim_configs}

We first explain the training of the proposed detector and compare its performance with MP \cite{raviteja2018interference}, UAMP \cite{yuan2020iterative}, BPIC \cite{Akosasih2021otfs}, and  EP \cite{FLONG2022EP-OTFS} OTFS detectors. We set $T=10$, $K=7$, $\Delta f=15$kHz, $k_{\rm max} = 3$, SNR=$15$dB, and employ $4$-QAM modulation. The carrier frequency is set to $f_c = 10$ GHz.
  
% In this section, we evaluate the performance of our detector by comparing the BER of our proposed detector with that of the MMSE OTFS \cite{singh2020low}, MP OTFS \cite{raviteja2018interference},
% AMP OTFS\cite{khumalo2016fixed}, and  UTAMP OTFS \cite{yuan2020iterative} detectors. 
% We set $L=12$, $K=7$,  and $\Delta f=15$kHz. In the channel, the delay index is $l_i \in [1,6]$ excluding the first path ($l_i = 0$), the Doppler shift index of $i$-th path, $k_i$, is uniformly drawn from $[-1,1]$ or $[-3,3]$ and the path gain $h_i$ is independently drawn from the complex Gaussian distribution $\mathcal{N}(0, 1/P)$. The $4$-QAM modulation is employed for the simulations. The number of of the transmitted QAM symbols per OTFS frame is $KL=84$ and the number of samples in CP is $N_{\rm CP}=6$.
 
 %\red{The number of the transmitted QAM symbols, sent in an OTFS frame, equals to the number of the paths where the symbols are randomly placed in the DD grids.}

%We consider both the low and high Doppler scenario where the Doppler frequency shift index ranges are $[-1, 1]$ and $[-3, 3]$, respectively. We consider delay index $[0, 6]$ with $4$-QAM modulation. 
% to the relative velocity of $506.25$ km/h, considered in $5$G applications\cite{surabhi2019diversity} and $[-3, 3]$, respectively} . We consider delay index $[0, 6]$, $4$-QAM modulation. 

\subsection{Training of the BPICNet OTFS Detector}

We implemented the BPICNet in PyTorch. The training was divided into $500$ epochs. In each epoch, $40$ batches of $256$ samples were generated, where a sample includes realizations of $\qx_{\rm DD}$, $\tilde{\qw}$, and $\qH_{\rm eff}$ satisfying \eqref{sysddmod-y=hx+w}. In each batch, we randomly chose $P\in \{6,\dots,14\}$ while the SNR values are uniformly distributed in a certain range. 
% The value of $P$ for samples in each batch are the same but they vary over batches.
Therefore, the BPICNet OTFS detector can be implemented in systems with dynamic changes in $P$  without retraining. 
% This is in contrast to the state-of-the-art OTFS NN detectors \cite{Naikoti2021VTC,Enku2021CNN}. 
We used Adam optimizer with  learning rate of $0.0001$ and applied the learning rate scheduler (PyTorch ReduceLROnPlateau) to adjust the learning rate based on the validation loss, computed
using additional $5000$ samples in each epoch. In each batch, we calculated the average mean square error loss as
\begin{equation}\label{Loss_func}
Loss= \frac{1}{256}  \sum_{w=1}^{256} \sum_{q=1}^{KL} \left( x_{q,w} - \hat{x}_{q,w}^{(T)}\right)^2,
\end{equation}
where $w$ is the index of the training samples in each batch and $\hat{x}_{q,w}^{(T)}$ refers to the final BPICNet symbol estimate of ${x}_{q,w}$ obtained from the last $T$-th layer.
% \blue{Note that the loss function can be calculated based on symbol estimates from all layers which will improve the convergence of the detector at the cost of performance loss.}
The NN parameters i.e., $\theta_1^{(t)},\theta_2^{(t)},\text{and } \theta_3^{(t)}, t = 1, \dots, T$ were adjusted to minimize the loss  in \eqref{Loss_func} using the  backpropagation.

\subsection{SER Comparison}

% \newcolumntype{L}[1]{>{\raggedright\let\newline\\\arraybackslash\hspace{0pt}}m{#1}}
% \newcolumntype{C}[1]{>{\centering\let\newline\\\arraybackslash\hspace{0pt}}m{#1}}
% \newcolumntype{R}[1]{>{\raggedleft\let\newline\\\arraybackslash\hspace{0pt}}m{#1}}

% We first consider two numbers of paths $P=6$ and $P=10$. The simulation results are shown in the Figs. \ref{Fig:IB-PIC-DSC-Result-good} and  \ref{Fig:IB-PIC-DSC-Result-good-highscattering}. Here, the UTAMP OTFS detector fails to achieve an acceptable detection performance. This is because the unitary transformation is not applicable in our system model. Additionally, the MP OTFS detector outperforms the AMP and MMSE detectors. Nevertheless, the B-PIC-DSC OTFS detector can achieve the lowest BER in comparison to the other counterparts.
Fig. \ref{Fig:SER}a demonstrates that the BPICNet OTFS detector significantly outperforms the other detectors except for the EP OTFS  detector when the number of moving  reflectors increases. The BPICNet achieves a similar SER performance to the EP with much lower complexity, as discussed in Section \ref{Complexity}. 
{Fig. \ref{Fig:SER}b  evaluates SER with respect to the OTFS matrix size. It is done by fixing $K=7$ while varying $L\in\{12,16,24,32\}$. The result shows the outperformance of the proposed detector compared to the state-of-the-art. The BPICNet detector has a similar convergence rate as the standard BPIC and UAMP detectors, as illustrated in Fig. \ref{Fig:SER}c. Nevertheless, the BPICNet assures a better SER performance}

% Fig. \ref{Fig:SER}b evaluates the SER versus maximum velocity of the moving reflectors, $v_{\rm max} =\frac{c k_{\rm max} \Delta f}{f_c K}$, where $c$ is the speed of light. 
% Similarly to the observation in Fig. \ref{Fig:SER}a, the BPICNet OTFS detector achieves the closest SER performance to the EP OTFS detector compared to the other state-of-the-arts. 

% \begin{figure}
% \centering
% \subfloat[P=14]
% {\includegraphics[width=0.51\textwidth]{LTE_Min_UESuFast_P14}\hfill}
% \subfloat[P=18]
% {\includegraphics[width=0.51\textwidth]{LTE_Min_UESuFast_P18}\hfill}
% \caption{A large number of mobile reflectors with $k_{max}=3$}
% \label{Fig:IB-PIC-DSC-Result-bad-highscattering}
% \end{figure} 

\section{Conclusion}
We proposed a BPICNet detector for an OTFS system. It achieved a high detection performance under rich scattering environment.
% , where there was a large number of moving reflectors. 
Our simulation results showed that the BPICNet achieved a similar EP performance.

\section*{Acknowledgment}
This research was supported by the research training program stipend from The University of Sydney,  Australian Research Council grant number DP210103410, and the  ARC Laureate Fellowship grant number FL160100032.

{\renewcommand{\baselinestretch}{1.1}
\begin{footnotesize}
\bibliographystyle{IEEEtran}
\bibliography{IEEEabrv,myBib}
\end{footnotesize}}

\end{document}

%% file: macro.TeX
    \def\Complex{{\rm\rule[.23ex]{.03em}{1.1ex}\kern-.3em{C}}}

    \newcommand{\be}{\begin{equation}} \newcommand{\ee}{\end{equation}}
    \newcommand{\bea}{\begin{eqnarray}} \newcommand{\eea}{\end{eqnarray}}
    \newcommand{\benum}{\begin{enumerate}} \newcommand{\eenum}{\end{enumerate}}

    \newcommand{\qa}{{\bf a}}

        \newcommand{\qh}{{\bf h}}

        \newcommand{\qr}{{\bf r}}
        \newcommand{\qs}{{\bf s}}

        \newcommand{\qv}{{\bf v}}
        \newcommand{\qw}{{\bf w}}
        \newcommand{\qx}{{\bf x}}
        \newcommand{\qy}{{\bf y}}

        \newcommand{\qA}{{\bf A}}

        \newcommand{\qF}{{\bf F}}
        \newcommand{\qG}{{\bf G}}
        \newcommand{\qH}{{\bf H}}
        \newcommand{\qI}{{\bf I}}

        \newcommand{\qR}{{\bf R}}

        \newcommand{\qX}{{\bf X}}

        \newcommand{\qDelta}{{\boldsymbol \Delta}}

        \newcommand{\qSigma}{{\boldsymbol \Sigma}}

        \newcommand{\qmu}{{\boldsymbol \mu}}

        \newcommand{\Ex}{{\sf E}}

%% file: syst_mod.tex
\section{System Model}\label{syst_mod}

We consider an  OTFS system, illustrated in Fig. \ref{fig:sys-mod-general}. The transmitter and receiver are equipped with a single antenna. The channel state information (CSI) is assumed to be known in the receiver side. Note that the OTFS system can be developed by applying the inverse symplectic finite Fourier transform  (ISFFT) and SFFT blocks at the OFDM transmitter and receiver, respectively \cite{Hadani2017}. 
In the following, we explain the details of the OTFS transmitter, channel,  and receiver.  
%In the following, we use subscripts $(\cdot)_{\rm DD}$ and $(\cdot)_{\rm TF}$ to indicate that the variables are in DD and TF domains, respectively.

%% <Created-Terms>
% $M$-QAM, TF, DD, ISFFT
\subsection{OTFS Transmitter}

In the transmitter side, we map the information binary sequences into the $M$-ary quadrature amplitude modulation ($M$-QAM) symbols where the constellation is denoted as $\Omega$. The matrix of QAM symbols $\qX \in \mathbb{C}^{L \times K}$ are assigned in the DD domain discretized to an  $L \times  K$ grid, where $L$ and $K$ are the numbers of subcarriers and  OTFS subframes in an OTFS frame.
% (see Fig. \ref{fig:otfs-frame}a). 
The quantization steps of the delay and Doppler axes are $1/(L \Delta f)$ and $1/(K T_s)$, respectively. $\Delta f$ and $T_s$ are the subcarrier spacing and duration of an OTFS subframe, respectively.
% The delay and Doppler axes are respectively sampled by ${L \Delta f}$ and $K T_s$.
% , where $k=0,\cdots,K-1,$ and $l=0,\cdots,L-1$ are the indices of discretized delay and Doppler shifts, respectively. 
We transform the matrix of symbols $\qX$ from the DD domain into the time-frequency (TF) domain by using the ISFFT, as described in Fig. \ref{fig:sys-mod-general}. The ISFFT is performed by applying $L$-points DFT to the columns and $K$-points IDFT to the rows of $\qX$ \cite{2019Raviteja_practical} i.e., $\qF_L \qX \qF_K^{H}$, where $\qF_L \in \mathbb{C}^{L\times L}$ and $\qF_K^{H} \in \mathbb{C}^{K\times K}$ are the DFT and IDFT matrices\footnote{The $(p,q)$-th entries of the $N$-points DFT and its inverse are $(\frac{1}{\sqrt{N}}e^{-j2\pi(p-1)(q-1)/N})$ and $(\frac{1}{\sqrt{N}}e^{j2\pi(p-1)(q-1)/N})$, ${p,q=1,\cdots,N}$, respectively.}, respectively. 
The TF domain is discretized to an $L\times K$ grid with uniform intervals $\Delta f$ (Hz) and $T_s=1/\Delta f$ (seconds) in frequency and  time axes. Therefore, an OTFS frame occupies a bandwidth of $L\Delta f$ and a duration of $KT_s$.
% , as depicted in Fig. \ref{fig:otfs-frame}a.
A Heisenberg transform, consisting of an $L$-points IDFT and pulse shaping waveform, is then applied to generate the time-domain transmitted signal $\qG_{\rm tx}\qF^{H}_L     (\qF_L\qX \qF_K^{H})  $, where   $\qG_{\rm tx} =  \qI_L$ for a rectangular waveform \cite{2019Raviteja_practical} with a duration of $T_s$.
% $\qG_{\rm tx} = {\sf diag}\left[g_{\rm tx}(0),g_{\rm tx}(T_s/L),\cdots,g_{\rm tx}((L-1)T_s/L)  \right] \in \mathbb{C}^{L\times L}$, $\sf diag[\cdot]$ denotes the operation to diagonalize a vector, $g_{\rm tx}(t)$ is a rectangular waveform. Note that for a rectangular waveform, $\qG_{\rm tx}$ reduces to the identity matrix $\qI_L$.
The transmitted signal is rewritten in a vector form as 
\begin{equation}
\qs = {\sf vec}(\qI_L \qX \qF_K^{H})  = (\qF^{H}_K\otimes \qI_L)\qx, \quad \qx  = {\sf vec}(\qX),
\label{eq:sysddmod-tx-heisenberg}
\end{equation}
% \begin{align}
% \qs &= 
% % {\sf vec}(\qG_{\rm tx}\qF^{H}_L\qX_{\rm TF}) \notag \\ 
% % &={\sf vec}(\qG_{\rm tx}\qF^{H}_L    \qF_L\qX_{\rm DD}\qF_K^{H})  \notag \\ 
% & = {\sf vec}(\qG_{\rm tx}\qX \qF_K^{H})  = (\qF^{H}_K\otimes \qG_{\rm tx})\qx
% \label{eq:sysddmod-tx-heisenberg}
% \end{align}
obtained  by using the Kronecker product rule\footnote{A matrix multiplication is often expressed by using vectorization with the Kronecker product. That is, ${\sf vec}(ABC) = (C^\top \otimes A){\sf vec}(B)$}.
 We consider the OTFS model as in \cite{2019Raviteja_practical}, where the cyclic prefix (CP) is  inserted only at the beginning of the OTFS frame.
For each OTFS frame, the time duration after adding CP is $KT_s + N_{\rm cp}\frac{T_s}{L}$, where $ N_{\rm cp}$ is equal to the index of the maximum delay.
% , as shown in Fig. \ref{fig:otfs-frame}b.

%, where each OTFS frame contains of $K$ number of OTFS subframes and the CP.

%We can then rewrite \eqref{eq:sysddmod-tx-heisenberg} in vector form as
%\begin{equation}
%\qs = {\sf vec}(\qS) = (\qF^{H}_K\otimes \qG_{\rm tx})\qx_{\rm DD},
%\label{eq:sysddmod-tx-vec}
%\end{equation}

\subsection{OTFS Wireless Channel}

The OTFS wireless channel is a time-varying multipath channel, represented by the impulse responses in the DD domain,  $ h(\tau, v) = \sum_{i=1}^P h_i \delta(\tau - \tau_i)\delta(v - \upsilon_i) $, where $\delta(\cdot)$ is the Dirac delta function, $h_i \sim \mathcal{N}(0, 1/P)$ denotes the $i$-th path gain,  $P$ is the number of propagation paths, and each path is associated with a moving reflector as described in Fig. \ref{fig:friendly}.
% \blue{where $h_i$, $\tau_i$, ${\upsilon_i}$ are the path gain, delay and Doppler shift on $i$-th path} 
The paths have different delay ($\tau_i$) and/or Doppler (${\upsilon_i}$) characteristics.
% ,  for example $ (\tau_1,v_1) = (1\text{ms},100\text{Hz})$ and  $(\tau_2,v_2) = (1\text{ms},50\text{Hz}) $.
% \gray{Each of the paths represents a channel between a moving reflector or transmitter and a receiver.} 
The delay and Doppler shifts are given as $ \tau_i = l_i\frac{T_s}{L} $ and {$\upsilon_i = (k_i+\kappa_i) \frac{\Delta f}{K}, $} respectively.  The integers $l_i \in [0,  l_{\rm max}]$ and $k_i \in [-k_{\rm max}, k_{\rm max}]$ denote the indices of the delay and Doppler shift in $i$-th path,  where $l_{\rm max} \leq L-1$ and $k_{\rm max} \leq \floor*{\frac{K}{2}}$ are the indices of the maximum delay and Doppler shift among all channel paths.
 {The fractional Doppler, $ -0.5< \kappa_i \leq 0.5$, $\kappa_i \in \mathbb{R}$, is set as $\kappa_i=0$. This means we do not consider fractional Doppler shifts in this work and leave this for future work.}
 % Nevertheless  the fractional Doppler shifts can be tackled by adding virtual integer taps in the delay–Doppler channel \cite{fractional_Doppler}. 
 The maximum number of paths in OTFS systems is $(2 k_{\rm max} +1 ) \times l_{\rm max}$. 

\subsection{OTFS Receiver}

% The received signal is obtained from sending the transmitted signal over the channel  $h(\tau,v)$, which includes the delay and Doppler terms. 
We discard the CP from a received OTFS frame and thus the time domain received signal is given as\cite{2019Raviteja_practical} 
% \begin{equation}
% r(n) =  \sum^P_{i} h_i e^{j2\pi \frac{k_i(n-l_i)}{KL}} s([n - l_i]_{KL}) + w(n),
% \label{eq:sysmod-r(n)-scalar}
% \end{equation}
% where $[\cdot]_{KL}$ denotes mod-$(KL)$ operation and $n=0,\cdots, KL - 1$. 
% We can rewrite \eqref{eq:sysmod-r(n)-scalar} in a matrix-vector form as
\begin{equation}
\qr = \qH\qs + \qw,
\label{eq:sysmod-r(n)}
\end{equation}
where $\qw$ is the independent and identically distributed (i.i.d.) white Gaussian noise that follows $\mathcal{N}(\mathbf{0}, \sigma^2 \qI_{KL})$, $\sigma^2$ is the variance of the noise, $\qH = \sum_{i=1}^P h_i \qI_{KL}(l_i) \qDelta({k_i})$, $\qI_{KL}(l_i)$ denotes a $KL\times KL$ matrix obtained by circularly left shifting the columns of the identity matrix by $l_i$, 
% , for example when $l_i =1$,
% \[
% \qI_{KL}(1) = \begin{bmatrix}
% 0 & \cdots & 0 & 1\\
% 1 & \ddots & 0 & 0\\
% \vdots & \ddots & \ddots & \vdots\\
% 0 & \cdots & 1 & 0\\
% \end{bmatrix}	.
% \]
 $ \qDelta({k_i}) = {\sf diag}\left[e^{\frac{j2\pi k_i(0)}{KL}}, e^{\frac{j2\pi k_i(1)}{KL}}, \cdots, e^{\frac{j2\pi k_i(KL - 1)}{KL}}\right]$ is a $KL \times KL$ Doppler shift diagonal matrix, and ${\sf diag}(\cdot)$ denotes a diagonalization operation on a vector.
%\end{equation}
%$
%\qDelta({k_i}) = \begin{bmatrix}
%e^{\frac{j2\pi k_i(0)}{KL}} & 0 & \cdots & 0\\
%0 & e^{\frac{j2\pi k_i(1)}{KL}} & \ddots & \vdots\\
%\vdots & \ddots & \ddots &  0 \\ 
%0 & \cdots & 0 & e^{\frac{j2\pi k_i(KL - 1)}{KL}}
%\end{bmatrix}	.
%$\\
% Note that the matrices $\qI_{KL}(l_i)$ and $\qDelta({k_i})$ describe the delay and Doppler shifts in \eqref{eq:sysmod-r(n)-scalar}, respectively.
% Let us now define a matrix  \[
% \qR \triangleq  \begin{bmatrix}
% r(0) & r(L) & \cdots & r((K-1)L)\\
% r(1)  & r(L+1) & \cdots & r((K-1)L + 1)\\
% \vdots & \vdots & \cdots &  \vdots \\ 
% r(L-1) & r(2L-1) & \cdots & r(KL-1)
% \end{bmatrix}	.
% \]
As shown in Fig. \ref{fig:sys-mod-general}, the received signal in the TF domain is obtained by applying the Wigner transform \cite{2019Raviteja_practical}, that consists of an $L$-points DFT and a pulse shaping waveform  with a duration of $T_s$. The TF domain received signal is $\qF_L\qG_{\rm rx}\qR$,
% \begin{equation}
% \qY_{\rm TF} = \qF_L\qG_{\rm rx}\qR,
% \label{eq:sysddmod-rx-wigner}
% \end{equation}
where 
% \gray{$\qG_{\rm rx} = {\sf diag}\left[g_{\rm rx}(0),g_{\rm rx}(/L),\cdots,g_{\rm rx}((L-1)/L)  \right] \in \mathbb{C}^{L\times L}$ and $g_{\rm rx}(t)$} 
$\qR = {\sf vec}^{-1}(\qr)$ and $\qG_{\rm rx}=\qI_L$ in the case of a rectangular waveform. 
 A symplectic finite Fourier transform (SFFT) is then performed by applying $L$-points IDFT to the columns and an $K$-points IDFT to the rows of  the TF domain received signal, expressed as $\qF^{H}_L (\qF_L  \qI_L \qR) \qF_K =  \qI_L\qR\qF_K$, which can be vectorized as 
%  to obtain the received signal in the DD domain, i.e.,
% \begin{align}
% \qY_{\rm DD}  &=\qF^{H}_L \qY_{\rm TF} \qF_K \notag \\
% &= \qF^{H}_L \qF_L\qG_{\rm rx}\qR \qF_K =  \qG_{\rm rx}\qR\qF_K.
% \label{eq:sysmod-rx-sfft}
% \end{align}
% By following the vectorization with Kronecker product rule, we can obtain
\begin{align} 
\qy&= {\sf vec}(\qI_L\qR\qF_K) 
% &= {\sf vec}(\qG_{\rm rx}\qR\qF_K) \notag \\
%&= (\qF_K^{\rm T} \otimes \qG_{\rm rx}){\sf vec}(R) \notag\\
%&= (\qF_K^{\rm T}  \otimes \qG_{\rm rx})\qr \notag \\
 = (\qF_K \otimes \qI_L)\qr.
\label{eq:sysmod-rx-vec}
\end{align}
Substituting \eqref{eq:sysddmod-tx-heisenberg} into \eqref{eq:sysmod-r(n)} and  \eqref{eq:sysmod-rx-vec} gives us 
\begin{align}
\qy = \qH_{\rm eff}\qx + \tilde{\qw},
%&= (\qF_K \otimes \qG_{\rm rx})H(\qF^{H}_K \otimes \qG_{\rm tx})\qx_{\rm DD} + (\qF_K \otimes \qG_{\rm rx})\qw \notag \\
%&= \qH_{\rm eff}\qx + \tilde{\qw},
\label{sysddmod-y=hx+w}
\end{align}
% $\qH_{\rm eff} = (\qF_N \otimes \qI_L)\qH(\qF^{H}_N \otimes \qI_L)$
% $\tilde{\qw} = (\qF_N \otimes \qI_L)\qw$ 
where $\qH_{\rm eff} = (\qF_K \otimes \qI_L)\qH(\qF^{H}_K \otimes \qI_L)$ and $\tilde{\qw} =(\qF_K \otimes \qI_L)\qw$ denote the effective channel and noise in the DD domain, respectively.  Note that $\tilde{\qw}$ is an i.i.d. Gaussian noise, since $\qF_K \otimes \qG_{\rm rx}$ is a unitary orthogonal matrix \cite{Hadani2017, 2019Raviteja_practical}.
The complex-valued model in \eqref{sysddmod-y=hx+w} can be transformed into an equivalent real-valued model, as explained in \cite{NIPS2009_839ab468}. To this end we consider the equivalent real-valued model of \eqref{sysddmod-y=hx+w} and keep the notations the same to make them uncluttered.